\documentclass[twocolumn,pra]{revtex4-1}
\usepackage{graphicx}
\usepackage{amsmath}
\usepackage{amssymb}
\usepackage{hyperref}
\usepackage{latexsym}
\usepackage{array}
\usepackage{times}
\usepackage{color}
\usepackage{subfigure}
\usepackage{setspace}
\usepackage{bm}

 

\begin{document}

\newcommand{\lp}{\ensuremath{\left(}}
\newcommand{\rp}{\ensuremath{\right)}}
\newcommand{\cred}{\color{red}}
\newcommand{\cblack}{\color{black}}
\newcommand{\e}[1]{\ensuremath{\times 10^{#1}}}

\title{Suppression of collisional shifts in a strongly interacting lattice clock}

\author{Matthew D. Swallows, Michael Bishof, Yige Lin, Sebastian Blatt, Michael J. Martin, Ana Maria Rey, and Jun Ye$\dagger$\\}
\affiliation{
\normalsize{JILA, National Institute of Standards and Technology and University of Colorado,}\\
\normalsize{Department of Physics,~University of Colorado}
\normalsize{Boulder, CO 80309-0440, USA}\\
\normalsize{$^\dagger$To whom correspondence should be addressed; }\\
\normalsize{E-mail:  Ye@jila.colorado.edu}
}


\begin{abstract}
Optical lattice clocks have the potential for extremely high frequency stability owing to the simultaneous interrogation of many atoms, but this precision may come at the cost of systematic inaccuracy due to atomic interactions.
Density--dependent frequency shifts can occur even in a clock that uses fermionic atoms if they are subject to inhomogeneous optical excitation~\cite{Campbell09, NISTYb}. Here we present a seemingly paradoxical solution to this problem. By dramatically increasing the strength of atomic interactions, we suppress collisional shifts in lattice sites containing $N>1$ atoms; strong interactions introduce an energy splitting into the system, and evolution into a many-particle state in which collisions occur is inhibited. We demonstrate the effectiveness of this approach with the JILA Sr lattice clock by reducing both the collisional frequency shift and its uncertainty by more than a factor of ten \cite{Ludlow08}, to the level of $10^{-17}$. This result eliminates the compromise between precision and accuracy in a many-particle system, since both will continue to improve as the particle number increases.
\end{abstract}

\maketitle

Strongly interacting quantum systems can exhibit counterintuitive behaviors, under both equilibrium and non-equilibrium conditions.
For example, in a multi-component Fermi degenerate gas, a frequency shift of a microwave transition remains finite close to a Feshbach resonance~\cite{Gupta2003,Zwerger07,Baym07}.
In low dimensions, the effective strength of atomic interactions can be significantly enhanced and as a result, particles tend to avoid each other to minimize their total energy. This tendency can lead to behavior that in many aspects resembles that of non-interacting systems. One such example is the Tonks--Girardeau regime of an ultracold Bose gas, in which the strong repulsion between particles mimics the Pauli exclusion principle, causing the bosons to behave like non-interacting fermions~\cite{Girardeau77,Weiss04,Paredes04,Innsbruck09}.
Here we describe how the enhancement of atomic interactions in a strongly interacting, effectively one-dimensional (1D) system suppresses collisional frequency shifts in an optical atomic clock, and we characterize the density-dependent shift to an unprecedented level of precision.

The $^{87}$Sr optical lattice clock at JILA has reached an overall fractional frequency uncertainty of $\sim$$1\times 10^{-16}$~\cite{Ludlow08,Swallows10}. This uncertainty is dominated by two contributions: atomic collisions and frequency shifts due to room--temperature blackbody radiation. The density-dependent shift arises from collisions between fermionic atoms that are subject to slightly inhomogeneous optical excitation~\cite{Campbell09,NISTYb}, and several theories of the underlying frequency shift mechanism have been proposed~\cite{Gibble09,Rey09,Copenhagen10}.
By tightly confining atoms in an array of quasi-1D potentials formed by a two-dimensional (2D) optical lattice, we significantly increase the strength of atomic interactions, to the point where the thermally averaged mean interaction energy per particle becomes the largest relevant energy scale.
Although our system does not meet the conventional definition of the strongly interacting regime (generally, that the interaction energy exceeds the system temperature), the interaction energy per particle is nevertheless larger than any other relevant dynamical energy scale, and thus the system is effectively strongly interacting. In this regime, collisions are suppressed because evolution into a many-particle state in which $s$-wave scattering can occur is energetically unfavorable.

Collisional frequency shifts could also be suppressed by confining atoms in a three-dimensional (3D) lattice with filling factor less than or equal to one per lattice site. However, vector and tensor shifts of the optical clock transition are a serious concern with a three-dimensional lattice clock using fermions~\cite{BoydPRA07}. A 3D lattice clock using bosonic $^{88}$Sr has been demonstrated~\cite{Katori08}, and its collisional shift was characterized at the level of $7\e{-16}$. In atoms without hyperfine structure, the $^1S_0\rightarrow{}^3P_0$ clock transition is completely forbidden, and therefore additional state mixing techniques must be used to enable it in bosonic Sr isotopes. These techniques result in sizable systematic shifts of the clock frequency that must be carefully controlled. The work presented here will allow operation of a fermionic lattice clock with a filling factor much greater than one and a greatly reduced sensitivity to collisional effects.

Our experiment employs ultracold fermionic $^{87}$Sr atoms that are nuclear spin-polarized (e.g., $I = 9/2$, $m_I=+9/2$). An ultranarrow optical clock transition, whose absolute frequency has been precisely measured \cite{Campbell08}, exists between the ground $^1S_0$ $\left(|g\rangle\right)$ and excited metastable $^3P_0$ $\left(|e\rangle\right)$ states. Atoms are trapped in a deep 2D optical lattice at the magic wavelength where the AC Stark shifts of $|g\rangle$ and $|e\rangle$ are matched~\cite{Ye08}. The 2D lattice provides strong confinement along two directions ($\hat X$ and $\hat Y$), and relatively weak confinement along the remaining dimension ($\hat Z$). Atoms in the lattice are sufficiently cold ($T_x \simeq T_y \simeq 2\ \mu$K) that they primarily occupy the ground state of the potentials along the tightly confined directions, with trap frequencies $\omega_X/2\pi$ $\sim$75-100 kHz and $\omega_Y/2\pi$ $\sim$45-65 kHz. This creates a 2D array of isolated tube-shaped potentials oriented along $\hat{Z}$, which have trap frequencies $\omega_Z/2\pi \sim$0.55-0.75 kHz.
We estimate that approximately 20\% of these lattice sites are occupied by more than one atom.
At a typical axial temperature $T_z$ of a few $\mu$K, various axial vibrational modes $n$ are populated in each tube. In a clock experiment the $|g\rangle \rightarrow |e\rangle$ transition is interrogated using Rabi spectroscopy. The atom--laser coupling is characterized by the bare Rabi frequency $\Omega^B_0$, which is defined in the absence of any motional effects. The optical frequency $\omega_L$ is detuned by an amount $\delta=\omega_L-\omega_0$ from the atomic resonance at $\omega_0$.  As described in Refs.~\cite{Campbell09, Blatt09}, any small projection of the probe beam along $\hat{Z}$ leads to a slightly different Rabi frequency $\Omega_n$ for each mode $\Omega_n(\eta_Z^2)$, where $\Omega_n<\Omega^B_0$. Here $\eta_Z= k_Z a_{ho}/\sqrt{2}$ is the Lamb-Dicke parameter, $a_{ho}= \sqrt{\hbar/(m_{\rm {Sr}}\omega_Z)}$ is the harmonic oscillator length, $m_{\rm {Sr}}$ is atomic mass, and $k_Z$ represents a small component of the probe laser wave vector along $\hat{Z}$, resulting in a typical $\eta_Z\sim$ 0.05.

To gain insight into the origin of the collisional frequency shift and the interaction-induced suppression, we consider a model system: two fermionic atoms, each of whose electronic degrees of freedom form a two-level, pseudo-spin 1/2 system ($|g\rangle$ and $|e\rangle$), confined in a 1D harmonic oscillator potential (a fully many-body treatment for an arbitrary number of atoms $N$ is discussed later in the text and developed in the Supplementary Information).
The internal degrees of freedom of these two identical fermions can be expressed using a collective-spin basis, comprised of three pseudospin-symmetric triplet states and an antisymmetric singlet state~\cite{Gibble09,Rey09}. Because the atoms are initially prepared in the same internal state $(|g\rangle)$, with their internal degrees of freedom symmetric with respect to exchange, the Pauli exclusion principle requires that their spatial wave function be antisymmetric. These atoms thus experience no $s$--wave interactions. If the atoms are coherently driven with the same Rabi frequency ($\bar{\Omega}$ = $(\Omega_{n_1} + \Omega_{n_2})/2 = \Omega_{n_1}$), their electronic degrees of freedom remain symmetric under exchange. Consequently, these atoms will not experience any $s$--wave interactions during the excitation of the clock transition. However, if $\Delta \Omega$ = ($|\Omega_{n_1}$ - $\Omega_{n_2}|$)/2 is not zero, the optical excitation inhomogeneity can transfer atoms with a certain probability to the antisymmetric spin state (singlet) that is separated from the triplet states by an energy $U$, since in this state atoms do interact. This interaction energy is what gives rise to a clock frequency shift during Rabi interrogation~\cite{Gibble09, Rey09}.

\begin{figure}
\begin{center}
 \includegraphics[width=\columnwidth]{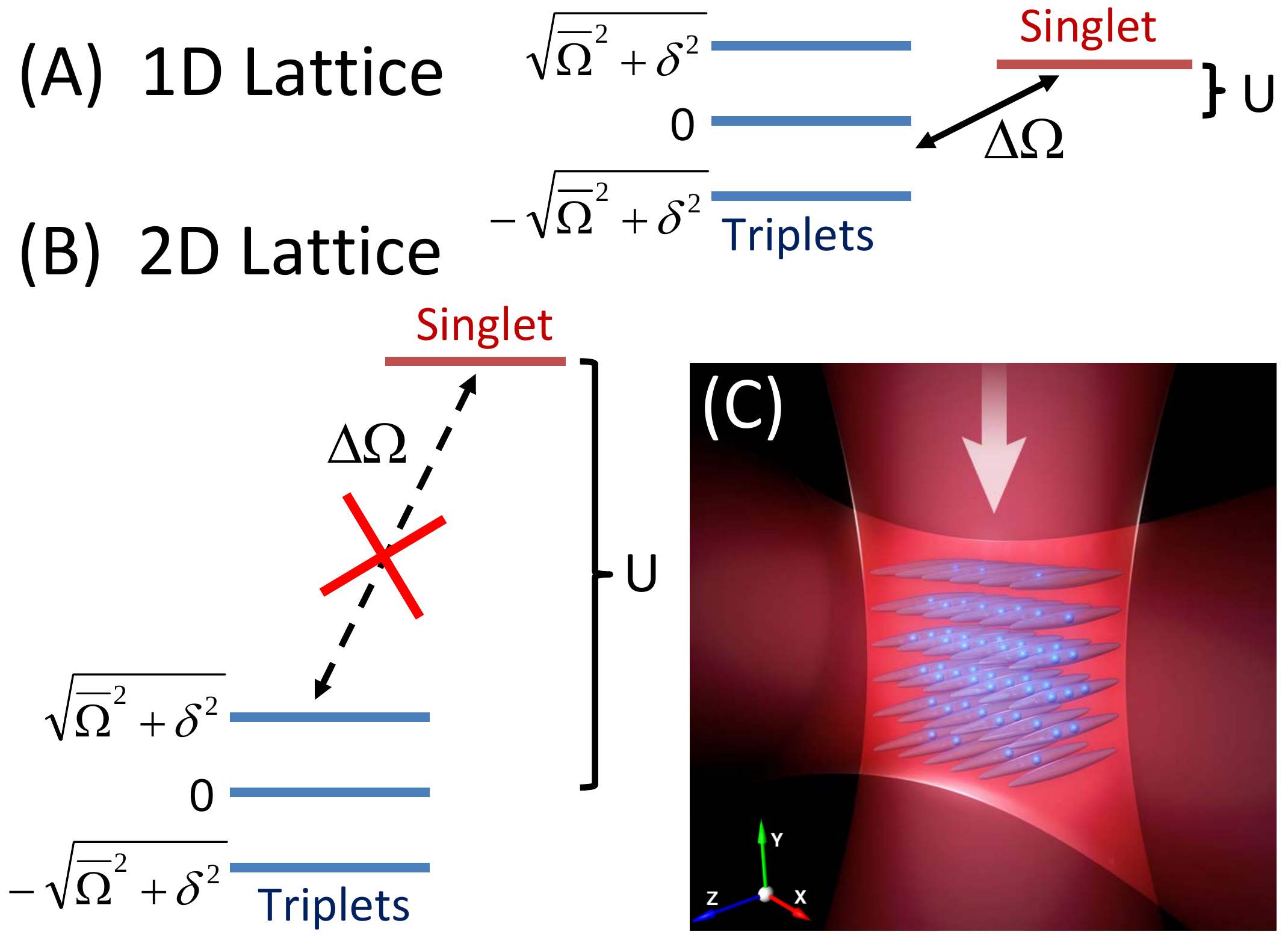}
 \end{center}
\caption{A schematic of the interaction blockade mechanism responsible for the suppression of collisional frequency shifts. (A) In prior experiments carried out in a 1D lattice, the atomic interaction is sufficiently weak that the energy of the singlet state lies within the energies of the dressed triplet states (which are distributed with an energy spread of the order of $\bar{\Omega}$). A weak excitation inhomogeneity characterized by $\Delta\Omega$ is capable of producing triplet-singlet mixtures, causing a collisional frequency shift proportional to the interaction strength $U$.  (B) In the 2D lattice, the interaction energy exceeds the atom-light Rabi frequency, creating an energy gap between the spin triplet and singlet states. A small excitation inhomogeneity cannot overcome the energy cost required to drive the transition. Evolution into the singlet state is inhibited and the collisional frequency shift is suppressed. (C) Quasi-1D tube-like optical potentials formed by two intersecting optical lattices. The laser which interrogates the clock transition propagates along $\hat Y$, the vertical axis.
} \label{fig1}
\end{figure}

Figure~\ref{fig1} contrasts the current 2D lattice experiment with prior studies carried out in a 1D lattice~\cite{Campbell09,NISTYb}. In a 1D lattice, the interaction energy $U$ is typically smaller than $2\bar{\Omega}$ (the energy spread of the driven triplet states at zero detuning). Consequently, any small excitation inhomogeneity $\Delta \Omega$ can efficiently populate the singlet state. By tightly confining atoms in a 2D lattice, one can reach the limit where $U \gg \bar{\Omega}$. In the presence of this large energy splitting, evolution into the singlet state is inhibited, and as a result the collisional frequency shift of the clock transition is suppressed. In this regime, the singlet state can only participate as a ``virtual" state in second-order excitation processes and the frequency shift scales as $\Delta \Omega^2/U$. In this limit, the energy-carrying singlet resonance has been shifted so far from the triplet resonances that it is completely resolved from them, and the remaining line-pulling effect can be negligible.
Such behavior is reminiscent of the dipolar blockade mechanism in a Rydberg atom gas~\cite{Saffman10}, where the interaction energy between an excited atom and its neighbors prevents their subsequent excitation.

The two-particle behavior described above can be generalized to an $N$-particle system ($N>2$).  The many-body Hamiltonian can be quantitatively modeled by defining a set of effective spin operators, $S^{x,y,z}_{n_j}$, in the $\left\{e,g\right\}$ basis. Here the subscript $n_j$ specifies the vibrational mode. The description of the system in terms of effective spin operators is valid provided those initially populated modes remain singly occupied by either a $|g\rangle$ or an $|e\rangle$  atom during the excitation process. The latter condition is satisfied in the Lamb-Dicke regime, $\eta_Z \ll 1 $, where one can neglect laser-induced inter-mode transitions.  To evaluate thermally averaged quantities, we restrict the calculations to a fixed set of initially populated modes $\vec{ n}= \{ n_1,\dots n_N\}$ and  sum over all possible mode configurations, weighted by the corresponding Boltzmann  factors. Under the rotating wave approximation, the Hamiltonian of the system becomes (see the Supplementary Information),

\begin{eqnarray}
\hat H^{S}_{\vec{ n}}/\hbar = &-&\delta {S}^z - \sum_{j=1}^N{\Omega_{n_j} {S}^{x}_{n_j}} \nonumber \\
&-&\sum_{j=1}^N\sum_{ j'\neq j }^N{\frac{ U_{n_j ,n_{j'}} }{2}(\vec{ S}_{n_j}\cdot \vec{ S}_{ n_{j'}} - 1/4)}. \label{many}
\end{eqnarray}
$S^{z,x} = \sum_{j=1}^N {S}^{z,x}_{n_j}$ are collective spin operators. The quantity $U_{n_j ,n_{j'}}= u I_{n_{j},n_{j'}}$ measures the strength of the interactions between two atoms in the antisymmetric electronic state. The interaction parameter $u = 4 \hbar a_{eg}^-/(m_{\rm{Sr}}V)$, where $a_{eg}^-$ is the singlet $g-e$ scattering length. Since the volume $V = \sqrt{\frac{\hbar}{m_{\rm{Sr}}\omega_X}\frac{\hbar}{m_{\rm{Sr}}\omega_Y}\frac{\hbar}{m_{\rm{Sr}}\omega_Z}}$, $u=4 \omega_{\perp} \frac{ a_{eg}^-}{a_{ho}}$,  where $\omega_{\perp}=\sqrt{\omega_X\omega_Y}$  is the mean transverse trapping frequency. $I_{{n_j},{n_j'}}$ is a mode overlap coefficient and its thermal average over all $n_j$ yields an interaction parameter $\langle U \rangle_{T_Z} \approx u \sqrt{ \frac{\pi}{40} \frac{\hbar \omega_Z}{k_B T_Z}}$. Note that  $u$ is directly proportional to $\omega_{\perp}$ and thus  increases with the transverse confinement.

The spin rotational invariance of the interaction term in $\hat H^S$ is key to understanding the basic physics~\cite{Zwierlein2003,Rey09}. Due to the rotational symmetry, the interaction term is diagonal in the collective spin basis $|S,M\rangle$, $S=0(\frac{1}{2}),\dots, N/2$ and $|M|\leq S$. For $N=2$, the spin basis is spanned by the triplet states $\left|S=1,M=\pm 1,0\right\rangle$ and the singlet $\left|S=0,M=0\right\rangle$. Among the collective states only the $S=N/2$ states are noninteracting. States with $S<N/2$ experience an interaction energy shift.

For a homogeneous excitation ($ \Omega_{\vec{n}}=\bar{\Omega}_{\vec{n}}$, where $\bar{\Omega}_{\vec{n}}=\sum_{n_j}\Omega_{n_j}/N$ is the mean Rabi frequency), the Hamiltonian commutes with $\hat{S}^2$ and thus $S$ is a conserved quantum number.  The eigenstates of the system are just the original collective spin states up to a rotation around the $y$ axis (for $N=2$ the eigenvalues are $\mp \sqrt{\bar{\Omega}_{\vec{n}}^2+\delta^2},0$ for $S=1$ and $U_{\vec{n}}$ for  $S=0$). If the system is prepared  in the $S=N/2$ manifold it will never experience any interaction effects, and there will be no collision-induced frequency shift.
In the presence of excitation inhomogeneity, $S$ is no longer conserved. During excitation of the clock transition, atoms will be transferred mainly between $S=N/2$ and $S=N/2-1$ states and will consequently experience a collisional frequency shift.  As demonstrated in the Supplementary Information, the collisional shift experienced by atoms in multiply occupied lattice sites ($N>2$) remains suppressed as $N$ increases. It also increases with
$\Delta{\Omega}_{\vec{n}}=\sqrt{\sum_{n}\Omega_{n}^2/N-\bar{\Omega}_{\vec{n}}^2}$.
We note that detuning inhomogeneity (for instance, atoms occupying different vibrational modes may have different resonance frequencies, $\delta$ = $\delta_{\vec{n}}$) has an effect similar to atom-laser coupling inhomogeneity, and the resulting shift can be modeled using Eq.~\ref{many}.

\begin{figure}
\begin{center}
\includegraphics[width=\columnwidth]{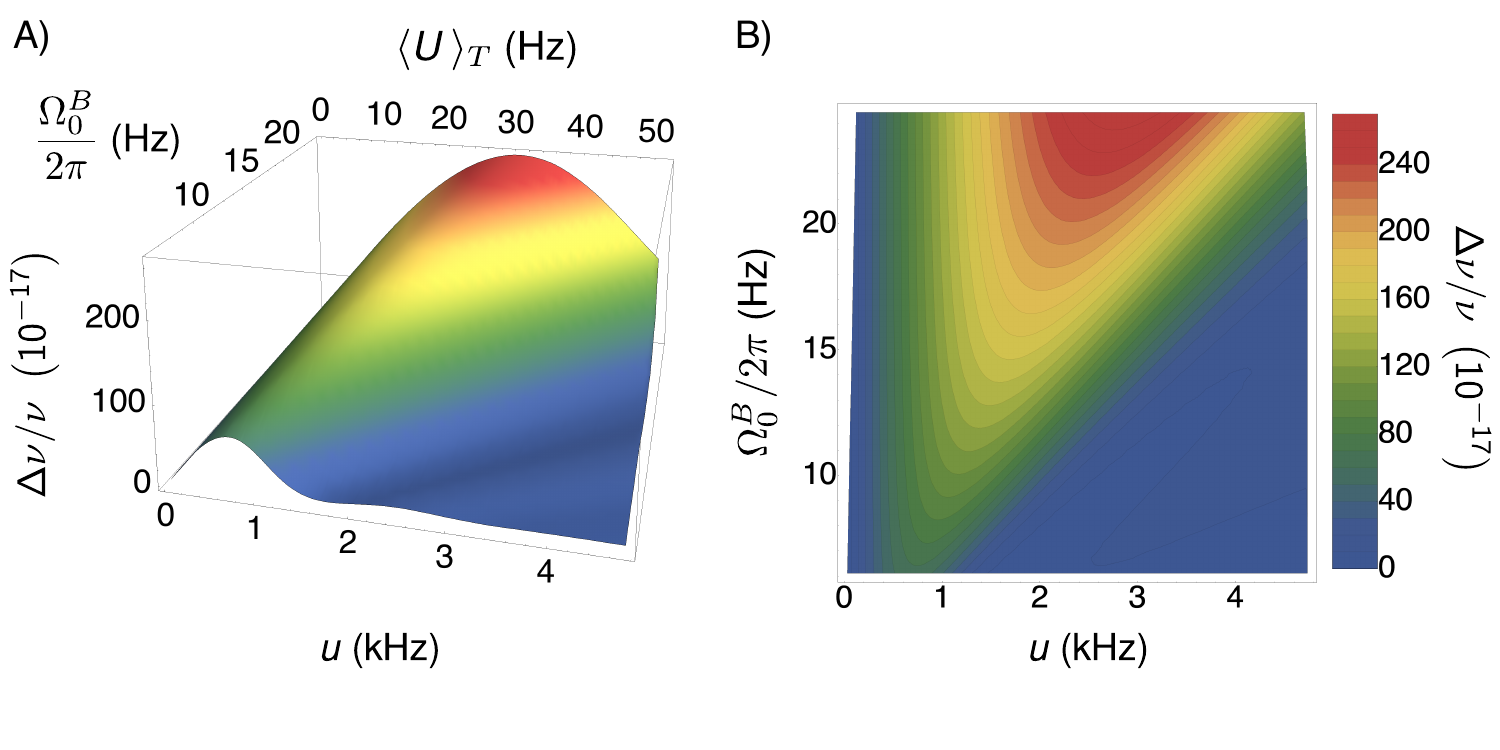}
\caption{Calculated suppression of the collision frequency shift (expressed as a fraction of the transition frequency) with sufficiently large atomic interactions. The criterion for suppression of the collision shift is $u \gg \Omega_0^B$. As $\Omega_0^B$ increases, a larger $u$ is required for clock shift suppression. Here the temperature along $\hat{Z}$ was set to  $T_{Z}$ = 6.5 $\mu$K, the axial Lamb-Dicke parameter to $\eta_Z=0.06$ and the laser detuning was fixed to achieve $30\%$ fractional population in the excited state $|e\rangle$.
The subfigures are A) a three-dimensional plot and B) a contour plot of the same theoretical function.
}
\label{fig2}
\end{center}
\end{figure}

In our clock experiments, the frequency of the laser which interrogates the $^1 S_0 \rightarrow$ $^3P_0$ transition (the clock laser) is modulated to probe the atomic resonance at two points $\delta_{1,2}$ on opposite sides of the line center. The mean frequency of the clock laser is steered to achieve equal population of the excited state at these two interrogation frequencies.
The collisional frequency shift $\Delta \nu $ is given by the change in the mean frequency as the  particle density is varied. We determine the transition lineshape by evaluating the thermal expectation value of the excited state population as a function of detuning (calculated from Eq.~\ref{many}), and compute the shift for a given excitation fraction as $\Delta \nu =(\delta_1+\delta_2)/2$.
A summary of these results is presented in Fig.~\ref{fig2}, which demonstrates the suppression of the collisional frequency shift as the interaction energy is increased.

A qualitative understanding of the suppression can be developed when we calculate the frequency shift by treating  $\Delta \Omega_{\vec{n}}$ as a perturbative parameter (see Supplementary Information for details). The perturbative analysis predicts, in the weak interacting regime of $u\ll \Omega_{0}^B$, a shift that scales linearly with $u$, $\Delta \nu = A(T_{Z},N) \eta_Z^4 N u$. Here $A(T_{Z},N)$ is a temperature-dependent coefficient with contributions from $\Delta \Omega_{\vec{n}}$, $\bar \Omega_{\vec{n}}$, and the mode-overlap coefficient $I_{n_j,n_{j'}}$. We also note that $\Delta \Omega_{\vec{n}}\propto \eta_Z^2$. This behavior is consistent with the standard mean-field expression of the density shift~\cite{Zwierlein2003,Gupta2003,Oktel1999, Oktel2002}. However, in the strong interacting regime of $u\gg \Omega_{0}^B$, the shift is suppressed as $\Delta \nu = B(T_{Z},N) \eta_Z^4 (\Omega_{0}^B)^2/(N u)$. Here $B(T_{Z},N)$ again includes the temperature-dependent effects. The suppression is consistent with the idea that in this regime the shift arises from a process that involves virtual occupations of non-fully symmetric states.

The suppression becomes less effective if ${\Omega_0^B}$ becomes comparable to $u$, or when $\Delta \Omega_{\vec{n}}$ increases at larger temperatures. These considerations  imply that   clock experiments based on Ramsey interrogation  will not easily satisfy the suppression conditions outlined here, since the short pulses applied in the  Ramsey scheme generally have a Rabi frequency  more than ten times larger than those used in Rabi spectroscopy.

To prepare the atomic system, we laser cool $^{87}$Sr atoms to about $2\ \mu$K inside a magneto-optic trap based on the weak $^1 S_0 \rightarrow\ ^3P_1$ transition (linewidth 7.6 kHz), and then load them into a 1D vertical lattice (along $\hat{Y}$) which overlaps with the MOT. The spatial distribution of occupied 1D lattice sites is determined by the vertical extent of the MOT cloud, which is approximately Gaussian with a standard deviation $\sigma_V=30\ \mathrm{\mu}$m. We then adiabatically ramp up the horizontal lattice (along $\hat{X}$) to load the atoms into a 2D lattice. The frequencies of the two lattice beams are offset from one another by 200~MHz to eliminate interference effects.
To remove any atoms trapped in the 1D vertical lattice outside of the 2D intersection region, we ramp the vertical lattice off and then back on again. The number of horizontal lattice sites occupied is then determined by the radial temperature of the vertical lattice along $\hat Y$.
100 ``rows" of tubes are approximately uniformly distributed along $\hat Y$, while the ``columns" distributed along $\hat X$ are loaded according to a Gaussian distribution with standard deviation $\sigma_H$ of 6-10 $\mathrm{\mu}$m.

After forming the 2D lattice, we perform Doppler and sideband cooling
using the $^1S_0\ \rightarrow\ ^3P_1,\ F=11/2$ transition.
Simultaneously, atoms are optically pumped to the $m_I=+9/2$ ground state sublevel, using $\sigma^+$-polarized light on the $^1S_0\ \rightarrow\ ^3P_1,\ F=9/2$ transition, directed along a bias magnetic field parallel to the $\hat Z$ axis. We determine the nuclear spin purity of the atomic sample to be greater than $97\%$ by scanning the probe laser over the clock transition frequencies for other nuclear spin states. We perform spectroscopy of the clock transition using a narrow linewidth laser propagating along $\hat{Y}$. The clock laser and both lattice beams are linearly polarized along $\hat{Z}$.

After cooling for about 30 ms, the sample temperature in the tightly-confined transverse dimension is lowered to 2 -- 2.5 $\mu$K.
We determine $T_Z$ by performing Doppler spectroscopy along $\hat{Z}$ and we vary $T_Z$ between 3 and $7\ \mu$K by applying additional Doppler cooling.
Trap frequencies along all three directions are determined via sideband spectroscopy and studies of parametric resonance.
Tunneling between lattice sites is suppressed because the potential gradient due to the dipole force lifts the degeneracy between lattice sites along both the $\hat X$ and $\hat Y$ directions. For the lattice with the lowest depth (along $\hat Y$) we estimate that the mean tunneling time is 150 ms for $T_X$ and $T_Y$ of $2.5\ \mu$K (details of this calculation are presented in SOM). Therefore, for our typical spectroscopy time of 80 ms, site-to-site tunneling can be neglected. We also note that we measure a frequency shift as a function of number of atoms loaded into the lattice, with all other lattice parameters kept constant, and thus our measurements are insensitive to effects that do not depend on $N$. We quantify the number of atoms loaded into the 2D lattice by detecting fluorescence on the strong  $^1 S_0\ \rightarrow\ ^{1}P_1$ transition at 461 nm.
With a total of $\sim4000$ atoms loaded into the 2D lattice, we estimate that 20--30\% of lattice sites are multiply occupied.

\begin{figure}
\begin{center}
\includegraphics[width=\columnwidth]{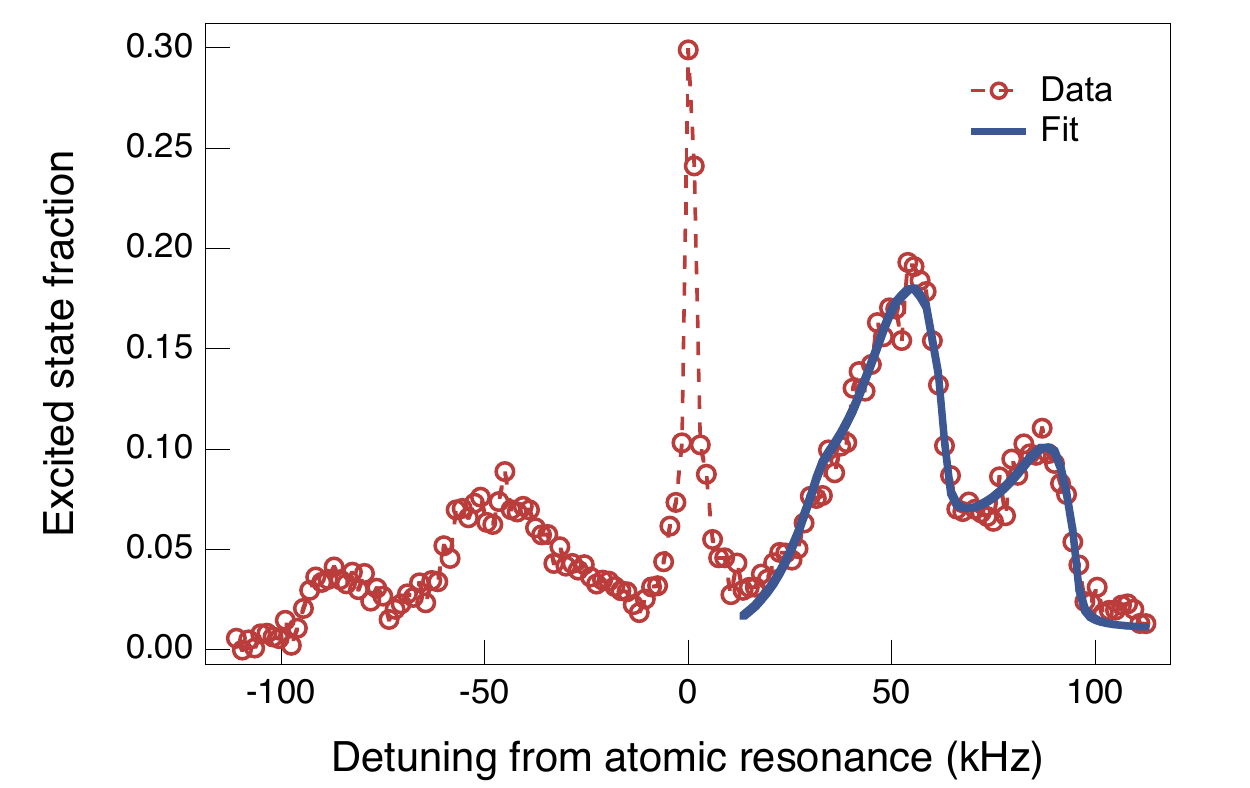}
\caption{Sideband spectroscopy of atoms confined in a 2D optical lattice. The clock laser power has been increased so that the carrier is power-broadened to $>1$ kHz and the sidebands are appreciably excited.
The two distinct sets of sidebands reflect trap frequencies in the tightly confined $X-Y$ plane. The features at detunings of $\pm90$ kHz are the first-order blue and red sidebands of the horizontal lattice. Due to experimental constraints, the horizonal lattice is not strictly perpendicular to $\hat Y$, and hence the clock laser (along $\hat Y$) also excites these sidebands.} \label{fig3}
\end{center}
\end{figure}

We build upon the method of extracting temperature information from vibrational sideband spectra that was derived for a 1D lattice~\cite{Blatt09}, and develop a model for the sideband features that we observe in a 2D lattice. These features can be understood in a fairly detailed fashion and our model reasonably describes the observed spectral features, as shown in Fig.~\ref{fig3}.
In the 1D lattice, all lattice sites are approximately equivalent, as the lattice's Rayleigh range is much larger than the spatial extent of the MOT cloud from which the lattice is loaded; in the 2D lattice this scenario no longer holds, as lattice sites near the wings of the lattice beams' Gaussian intensity profiles are significantly shallower than those near the center of the beam intersection region.
Only after taking the distribution of site depths into account does the axial temperature extracted from the sideband model become comparable with that determined from Doppler spectroscopy along $\hat{Z}$. The width of the sideband features is a result of the coupling between the longitudinal and transverse degrees of freedom in the lattice \cite{Blatt09}. In the 2D lattice, the sideband shape is also affected by the distribution of trap depths. The spread of trap frequencies accounts for the broadened sideband lineshape without requiring an exaggerated temperature along $\hat{Z}$. The sharper edges of the sidebands at the largest detunings from the carrier are due to atoms trapped near the center of the beam intersection region, where the trap depths are greatest.

Spectroscopy of the clock transition is performed with an 80-ms pulse, resulting in a Fourier-limited linewidth of $\sim10$ Hz. The clock laser is locked to the atomic resonance. The high--finesse Fabry--Perot cavity~\cite{Ludlow07} used to narrow the clock laser's linewidth is sufficiently stable over short time scales that it can be used as a frequency reference in a differential measurement scheme~\cite{Boyd07}. A single experimental cycle (e.g., cooling and trapping atoms, preparing the 2D lattice, and interrogating the clock transition) requires about 1.5 s, and we modulate the sample density every two cycles. The corresponding modulation of the atomic resonance frequency relative to the cavity reference is a measurement of the density shift.

When the sample density is varied, the spatial distribution of atoms in the 2D lattice might change. As lattice sites near the center and at the edge of the beam intersection region have different trap depths, this could allow AC Stark shifts to contaminate our measurements, unless the lattice beams are tuned to the magic wavelength~\cite{Ye08}. In the experiment we stabilize the two lattice frequencies to be $+100$ MHz and $-100$ MHz from the experimentally determined magic wavelength~\cite{Ludlow08} to minimize this effect. We set a limit on this potential systematic by introducing a large frequency offset between the two lattices; with a 2 GHz frequency offset no difference in the density shift was observed with an uncertainty of $1\e{-16}$, and we therefore estimate that with the 200 MHz offset any potential corruption of our collisional shift measurements is at or below the $1\e{-17}$ level.
The 200 MHz separation used for the data presented here was chosen for reasons of experimental convenience, and in a full implementation of a 2D lattice clock both beams could be tuned much closer to the magic wavelength to reduce the associated AC Stark shift.

\begin{figure}
    \begin{center}
\includegraphics[width=\columnwidth]{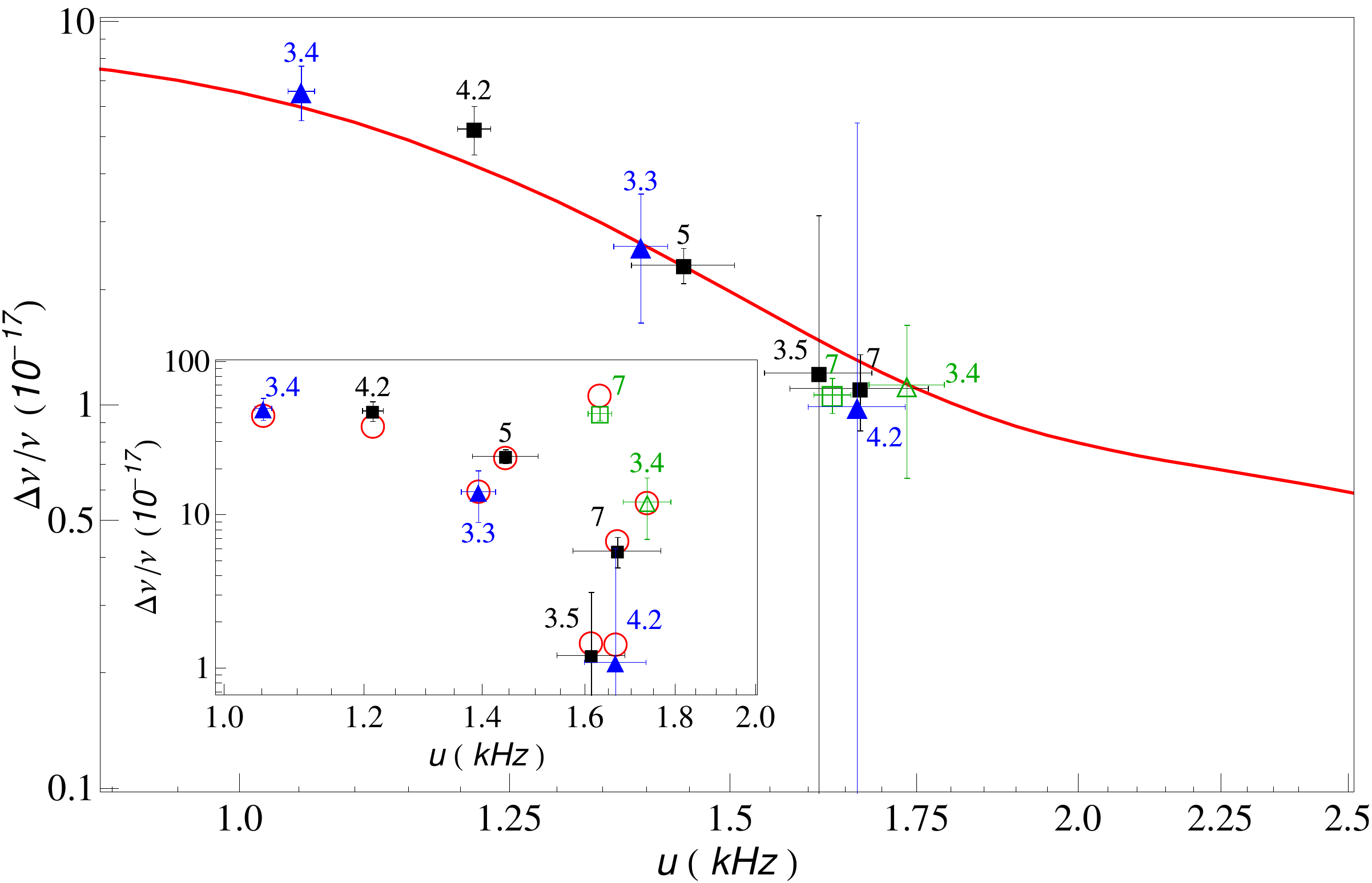}
\caption{Experimental observation of the suppression of the collisional frequency shift with increasing interaction energy
$u=4\omega_{\perp} \frac{ a_{eg}^-}{a_{ho}}$. For both the main figure and the inset, the x-axis is plotted under the assumption of $|a_{eg} ^-| = 40\ a_0$. We varied three important parameters: intensity of the horizontal lattice beam, axial temperature  and bare Rabi frequency.  In the inset we explicitly show the experimental data with $T_Z$  indicated on top of each point. The back and blue colored symbols were taken at $\Omega^B_0$ and the green colored symbols at $2\Omega^B_0$  and twice the interrogation time yielding a constant Rabi pulse area. We also use squares and triangles to distinguish two different sets of data points measured under slightly  different lattice  configurations  (different beam waist). In the inset the corresponding spin model predictions are also displayed (red open circles) using a singlet $g-e$ scattering length, $a_{eg} ^- =- (35$-$50)\ a_0$, and  $N=2$. Each theory point is calculated  using the  $T_Z$ and  $\Omega^B_0$  at which experimental data were taken, $\Delta\nu^T(\omega_Z^i,T_Z^i,u)$. The value of $u$ was varied by changing the horizontal lattice intensity, $I_X$, which also modified $\omega_Z$, $\omega_Y$, and the Lamb-Dicke parameter. The theoretical curves were scaled by the fraction of the atomic population in doubly occupied lattice sites. The variation of $\omega_Z$ and $u$ with $I_X$ was explicitly taken into account in theory which used $\eta_Z=0.046$ and $\omega_Z= 2\pi \times 0.7$ kHz at the  point  with the smallest collisional shift. The data confirm three trends in the prediction: the collisional shift $\Delta\nu$ decreases with increasing $u$ at similar temperature and trapping conditions, $\Delta\nu$ increases with increasing $\Omega^B_0$ at similar temperature and trapping conditions, and $\Delta\nu$ decreases with smaller $T_Z$. To overcome the difficulty of comparing the shift at different experimental conditions we rescaled the points by a factor of $\Delta\nu^T(\omega_Z^\mathrm{fix},T_Z^\mathrm{fix},u)/\Delta\nu^T(\omega_Z^i,T_Z^i,u)$ with $\omega_Z^\mathrm{fix}= 2\pi \times 0.7$ kHz and $T_z^\mathrm{fix}=3.5$ $\mu\mathrm{K}$. The main plot shows the good agreement between the theoretical curve of $\Delta\nu^T(\omega_Z^\mathrm{fix},T_Z^\mathrm{fix},u)$ in red solid line and rescaled experimental data.} \label{fig4}
\end{center}
\end{figure}

We perform measurements at several trap depths to directly observe the interaction-induced suppression of the collisional frequency shift.
To access different interaction energies, we vary the intensity of the horizontal lattice beam ($I_X$), which results in the change of mainly $\omega_X$ but also $\omega_Y$ and $\omega_Z$. The change in $\omega_Y$ arises from the fact that the laser beams that create the two lattices are not orthogonal but instead at an angle of $71^{\circ}$. The change in $\omega_Z$ is due to the Gaussian profile of the beams. Since $u \propto \sqrt{\omega_X \omega_Y \omega_Z}$, an increase of the horizontal beam power leads to a monotonic increase of $u$.  We observe a significant decrease of the collisional shift with increasing horizontal lattice power, as shown by the data points (filled black squares and blue triangles) in Fig.~\ref{fig4} (inset). Squares and triangles indicate data taken with slightly different beam waists. We have also studied the dependence of the collisional shift on the Rabi frequency used to drive the clock transition. $\Omega^B_0$ was increased by a factor of two, and the interrogation time was decreased by 2, yielding a constant Rabi pulse area.
Under these conditions, we observe that the collisional shift under similar temperature and trapping conditions increases sharply (green open square and green open triangle in Fig.~\ref{fig4} (inset)), confirming that the shift suppression mechanism will not operate effectively for short, higher Rabi frequency pulses.
This behavior is inconsistent with a frequency shift due to tunneling between lattice sites since points at similar trapping conditions exhibit substantially different shifts. Also shown in the inset of Fig.~\ref{fig4} is the shift predicted by the spin model, assuming two atoms per lattice site. The theoretical points are scaled by the fraction of the atomic population in doubly occupied lattice sites. The red open circles colors are the theory results, $\Delta\nu^T(\omega_Z^i,T_Z^i,u)$  with $i=1,\dots,9$, obtained at different temperatures and trapping frequencies  corresponding to the actual experiment conditions under which the data were taken. The data are consistent with the modeled shift, assuming $a_{eg}^-$ $= -(35$--$50)\ a_0$ (where $a_0$ is the Bohr radius),  $\eta_Z=0.046$, and $\omega_Z= 2\pi \times 0.7$ kHz at the  point  with the smallest collisional shift.  These parameters correspond to a value of $\langle U \rangle_{T_Z}/\langle \bar{\Omega} \rangle_{T_Z}= 9.1$ for this case. However, we note that $a_{eg}^-$ has not been measured independently, and therefore the interaction energy cannot be accurately estimated from the known trapping frequencies.

Since the temperature and trapping conditions substantially varied for different experimental data points, some scaling is required to make direct comparisons between data in Fig.~\ref{fig4} (inset) and the behavior predicted in Fig.~\ref{fig2}. To help visualization of the experimental confirmation of the interaction suppression mechanism, we rescaled the measured experimental values of the shift by a factor extracted from the theoretical model $\Delta\nu^T(\omega_Z^\mathrm{fix},T_Z^\mathrm{fix},u)/\Delta\nu^T(\omega_Z^i,T_Z^i,u)$. Fig.~\ref{fig4} shows that after rescaling all data points lie very close to the theoretical curve of fractional frequency shift  {\it vs.} $u$ at constant $\omega_Z^\mathrm{fix}= 2\pi \times 0.7$ kHz and $T_z^\mathrm{fix}$ = 3.5 $\mu$K.

The sign of the observed shift is negative, i.e., an increased sample density shifts the atomic resonance to lower frequencies.Previous studies of the collisional shift in a 1D optical lattice~\cite{Ludlow08,Campbell09} are consistent with this observation. The simple mean-field analysis used in~\cite{Blatt09} indicated a negative scattering length, but a more sophisticated many-body treatment~\cite{Rey09} showed that the experimental data were also consistent with a positive scattering length. From the present data set we can unambiguously conclude that $a_{eg} ^-$ is negative.

\begin{figure}
\begin{center}
\includegraphics[width=\columnwidth]{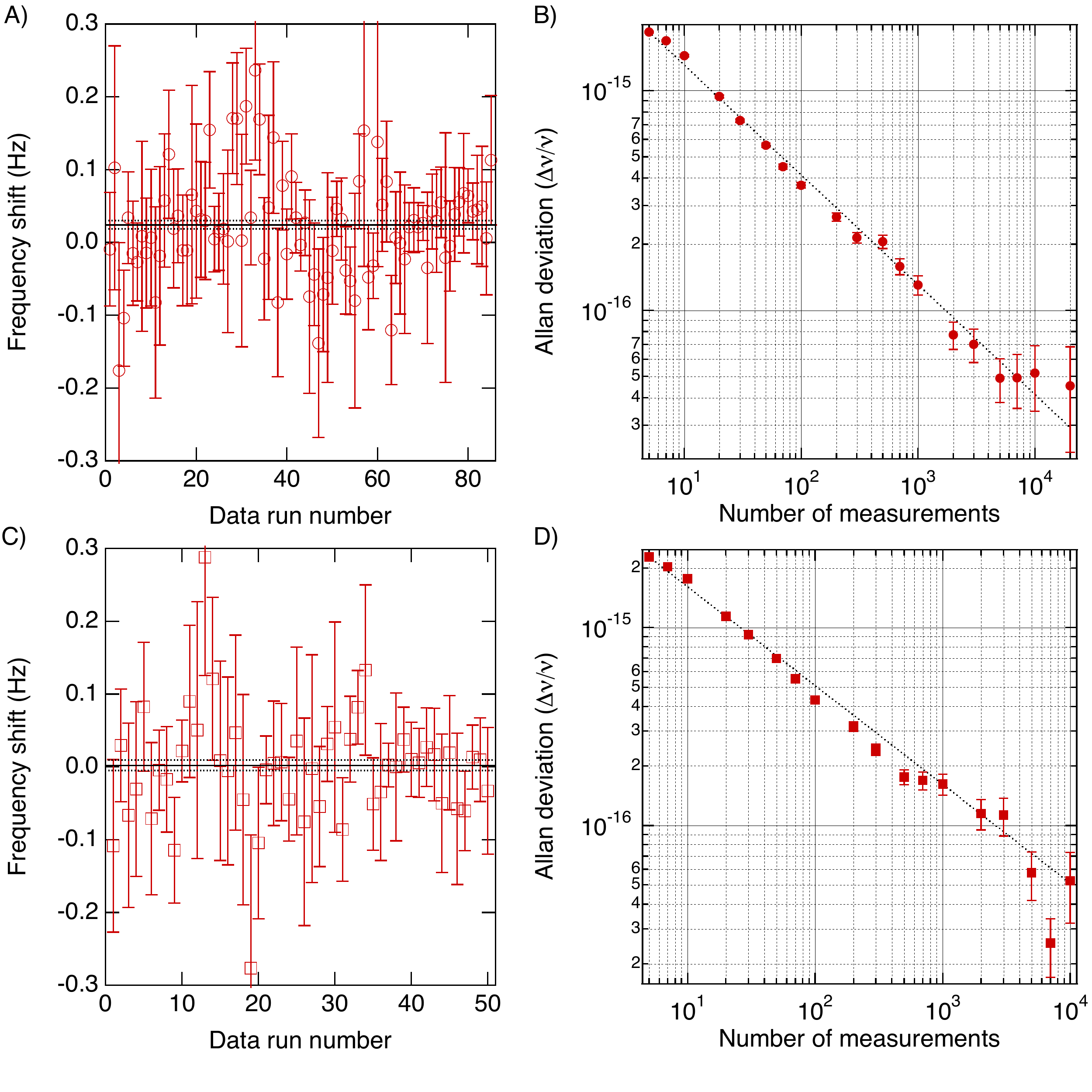}
\caption{Data records of collision-induced frequency shift measurements for $^{87}$Sr atoms confined in a 2D optical lattice. Each point represents a data set collected from a continuous operation of the Sr clock, with error bars determined from the standard error of that data set.
The weighted mean and weighted error of all the data are determined from the shift and error values of each data set, and the weighted error is scaled by the square-root of the reduced chi-square, $\sqrt{\chi_\mathrm{red}^2}$.
These are shown as the solid and dashed horizontal lines in panels (A) and (C).
Panels (B) and (D) show the corresponding Allan deviations (ignoring dead time between data runs) of the frequency shift records displayed in (A) and (C), respectively. Each measurement represents a differential comparison between two density conditions. Under typical clock operating conditions ($N \simeq 2000$), the weighted mean and the weighted standard error of the fractional frequency shift are $(5.6 \pm 1.3)\e{-17}$ at $T_Z$ = 7 $\mu$K ((A) and (B)) and $(0.5 \pm 1.7)\e{-17}$  at $T_Z$ = 3.5 $\mu$K ((C) and (D)). For the 7 $\mu$K data the reduced chi-square was $\sqrt{\chi_\mathrm{red}^2} = 0.84$, and for the $3.5\ \mu$K data $\sqrt{\chi_\mathrm{red}^2} = 0.73$.} \label{fig5}
\end{center}
\end{figure}

We have made an extensive series of collisional shift measurements at the largest trap depths available to us. The results of these measurements are displayed in Fig.~\ref{fig5}. The free-running clock laser has a stability of about $1.5\e{-15}$ at time scales of 1--10 seconds~\cite{Ludlow07}.
Therefore, a substantial integration time is required to determine the collisional shift with an uncertainty of $1\e{-17}$. Frequency drifts are minimized by measuring the long-term drift in the resonance frequency (relative to the ultrastable reference cavity) and applying a feed-forward correction to the clock laser. The correlation between the atomic resonance frequencies and the density of trapped atoms was
calculated by analyzing overlapping sequences of four consecutive measurements and eliminating frequency drifts of up to second order~\cite{Dress77}.
Approximately 60 hours of data were acquired at $T_Z$ = 7 $\mu$K over a $\sim 2$ month time period for the record shown in Fig.~\ref{fig5} (A).
Each data point represents a period during which the clock was continuously locked, with error bars determined from the standard error of the measurements in that data set.
The error bars of each data point were scaled by a correction factor $f_{cor}\simeq1.79$ that accounts for the fact that, due to the overlapping string analysis, each individual four-measurement sequence is only partially independent of its neighbors (see SOM).
The weighted mean and weighted standard error of these data determine the value of the shift.
On longer time scales, the data are underscattered, and we therefore scale the final error value by the square root of the reduced chi-square of the data set.
At an axial temperature $T_Z \simeq 7\ \mu$K, the collisional shift in our 2D lattice clock was measured to be $(5.6 \pm 1.3)\e{-17}$ in fractional units, with $\sqrt{\chi^2_{red}}\simeq0.84$. At a lower $T_Z$ of 3.5 $\mu$K, the collisional shift is reduced to $(0.5 \pm 1.7)\e{-17}$,  with $\sqrt{\chi^2_{red}}\simeq0.73$ (the record in Fig.~\ref{fig5} (C)). The corresponding Allan deviations of both data sets are shown in subpanels (B) and (D) of Fig.~\ref{fig5}. The data in both Fig.~\ref{fig4} and Fig.~\ref{fig5} have been analyzed using four different methods, the details of which are given in the SOM. The results of all analyses are consistent with each other to within the stated error bars.

Figures~\ref{fig4} and~\ref{fig5} demonstrate the suppression of the clock frequency shift as the system approaches the strongly interacting regime. We note that, relative to previous measurements of collisional shifts in a 1D optical lattice~\cite{Ludlow08, Campbell09}, the atomic density in a 2D lattice is much higher. After accounting for the increased temperature in our current experiments, we find that the local density in a lattice site occupied by two particles is $\sim 1.4\e{13}\ \mathrm{cm}^{-3}$, an order of magnitude larger than the average density in earlier 1D lattice experiments. Therefore, given a similar level of excitation inhomogeneity, if the collisional shift in the 2D lattice were not suppressed, we would expect a significantly larger shift than that in the 1D lattice experiments, even after assuming that only $20-30\%$ of lattice sites are contributing.

The results presented here demonstrate that detailed understanding of a quantum system can result in dramatic improvements in the areas of precision measurement and atomic clocks. This advance removes an important obstacle to further increasing the precision and accuracy of neutral atom-based optical clocks. Increasing the number of atoms loaded into our 2D lattice system will enable us to improve the stability of our clock without imposing an onerous systematic effect.
As clock lasers become more stable, we will be able to increase the duration of the Rabi interrogation pulse, thus decreasing the Rabi frequency. This will allow optical lattice clocks to operate in  the regime where the density shift is fully suppressed ($\sim1/u$) and further reduce the collisional shift systematic well into the 10$^{-18}$ domain. This, together with the fact that in the strongly-interacting regime the collisional shift will remain suppressed as more atoms are loaded into individual lattice sites, will enable neutral atom clocks to operate with the large sample sizes needed to achieve the highest possible stability.

\textbf{Acknowledgements}
We thank C. Benko for technical assistance and A. D. Ludlow and A. Gorshkov for useful discussions. This work was supported by a grant from the Army Research Office with funding from the DARPA OLE program, the US National Institute of Standards and Technology, the US National Science Foundation (NSF) Physics Frontier Center at JILA, and the Air Force Office for Scientific Research. M.D.S. is supported by a National Research Council Postdoctoral Fellowship. M.B. is partially supported by an NDSEG Graduate Fellowship. Y.L.'s permanent address is National Institute of Metrology, Beijing, China.

\clearpage

\appendix*

\section{Supplementary Online Material}
\subsection{Many-body Hamiltonian for spin--polarized fermionic atoms}

Here we will consider a spin polarized ensemble of fermionic atoms (e.g.  $I =9/2$) with two accessible   electronic degrees of freedom associated to the ${^1S_0}(g)-{^3P_0}(e)$ states.  We focus on the case where the atoms  are  trapped in an external potential $V(\mathbf{R})$ that is the same for $g$ and $e$ (i.e.~at the ``magic wavelength" \cite{Ye2008}). If the atoms are  illuminated by a linearly polarized laser beam with bare Rabi frequency $\Omega^B_0$ they are governed by the following
many-body Hamiltonian \cite{Gorshkov2009,Gorshkov20092,Rey2010,Hermele2009}
\begin{eqnarray}
\hat H &=& \sum_{\alpha }  \! \int  \!\! d^3  \mathbf{R} \hat \Psi^\dagger_{\alpha }   \left(- \frac{\hbar^2}{2 m_{\rm Sr}} \nabla^2 + V(\mathbf{R})\right) \hat \Psi_{\alpha } \nonumber \\  
&+& \hbar u_{eg}^-  \! \int \!\! d^3 \mathbf{R} \hat \rho_e \hat \rho_g +\hbar \omega_0  \! \int  \!\! d^3 \mathbf{R} \hat \rho_e \nonumber  \\
&-& \frac{ \hbar \Omega_0^B}{2}  \!
\int  \!\!  d^3 \mathbf{R} (\hat \Psi^\dagger_{e } e^{-i (\omega_L t- \mathbf{k} \cdot \mathbf{ R})} \hat \Psi_{g } + {\rm h.c.}).
\label{ham0}
\end{eqnarray}
Here $\hat \Psi_{\alpha }(\mathbf{R})$ is a fermionic field operator at position $\mathbf{R}$ for atoms with mass $m_{\rm Sr}$  in electronic  state  $\alpha = g$ ($^1S_0$) or $e$ ($^3P_0$), while $\hat \rho_{\alpha }(\mathbf{R}) =
\hat \Psi^\dagger_{\alpha }(\mathbf{R}) \hat \Psi_{\alpha }(\mathbf{R})$ is the corresponding density operator.
 Since nuclear spin polarized fermions are in a symmetric nuclear spin state, their  $s$-wave interactions are characterized by only one scattering
length $a_{eg}^-$, with the corresponding interaction parameter $u_{eg}^-= 4 \pi \hbar a_{eg}^-/m_{\rm Sr} $, describing collisions between two atoms in the antisymmetric electronic state.  The laser with frequency $\omega_L$ and wavevector $\mathbf{k}$ is detuned from the atom transition frequency $\omega_0$ by $\delta=\omega_L-\omega_0$.

We consider the situation in which a deep 2D lattice   freezes the atomic motion in the transverse $X-Y$ plane
creating an array of one dimensional tubes. Along the longitudinal $Z$-direction the net effect of the lattice is to induce a weak harmonic confinement with frequency $\omega_Z$. While the deep 2D lattice confines  the atoms to the lowest  vibrational mode, we  allow mobility along the longitudinal direction. It is then convenient to expand the field operator, $\hat{\Psi}_{\alpha}(\vec{R})$, in a harmonic oscillator basis,
$\hat{\Psi}_{\alpha}(\mathbf{R})=\phi^X_0(X) \phi^Y_0(Y)  \sum_{ n} \hat{c}_{\alpha n} \phi_{n} (Z) $, where
$\phi_0^{X,Y}$ and $\phi_n$ are, respectively, the transverse and the longitudinal harmonic oscillator eigenmodes and $\hat{c}^\dagger_{\alpha  n}$ creates a fermion in mode $n$ and electronic level $\alpha$.

Following Refs.~\cite{Campbell2009, Blatt2009,Rey2010,Gibble2009}, we assume that the probe is slightly misaligned with a small component along the  $Z$-direction: $\mathbf{k} = k_Y \hat Y + k_Z \hat Z$ with $|k_Z/k_Y| \ll 1$. Defining  $\Omega_{n,n'}= \Omega_0 e^{-(\eta_Y^2)/2} L_0(\eta_Y^2)  \langle \phi_{n}(Z)|e^{ i k_Z Z}|\phi_{n'}(Z)\rangle$, where $\eta_{Y,Z}=k_{Y,Z}a_{ho}^{Y,Z}/ \sqrt{2}\ll 1$ are the Lamb-Dicke parameters, $a_{ho}^{Y,Z}=\sqrt{\frac{\hbar}{m_\mathrm{Sr}\omega_{Y,Z}}}$ the corresponding harmonic oscillator lengths  and $L_n$ are Laguerre polynomials \cite{Wineland1979}, laser induced  sideband  transitions can be neglected if $ \Omega_{n,n' \neq n}\ll \omega_Z$ and $\delta\ll\omega_Z$. These conditions are satisfied in the experiment that operates in the resolved sideband regime, where the clock probe Fourier-limited spectral resolution is 10 Hz, $\ll \omega_z/2\pi$ (700 Hz). Hence, $\hat H$  can be rewritten in the rotating frame as 
\begin{widetext}
\begin{equation}
\hat H=- \hbar \delta \sum_{n=0}^{\infty} \hat{n}_{e n} +\sum_{\alpha} \sum_{n=0}^{\infty} E_{n} \hat{n}_{\alpha n} 
-\sum_{n=0}^{\infty}{\frac{\hbar \Omega_{n}}{2}(\hat{c}^\dag_{g n}\hat{c}_{e n}+ \rm{h.c})} 
+\frac{\hbar u}{2}\sum_{n=0}^{\infty}\sum_{n'=0}^{\infty}\sum_{n''=0}^{\infty} \sum_{n'''=0}^{\infty}{I_{n n' n'' n'''}\hat{c}^\dag_{e n}\hat{c}_{e n'}\hat{c}^\dag_{g n''}\hat{c}_{g n'''}},
\label{manyzero}
\end{equation}
\end{widetext}
where
\begin{equation}
u=4  \omega_{\perp} \frac{ a_{eg}^-}{a_{ho}} \nonumber
\end{equation}
and 
\begin{equation}
I_{n n' n'' n'''}=\frac{\int e^{-2\xi^2 }H_{n}(\xi) H_{n'}(\xi) H_{n''}(\xi)H_{n'''}(\xi)d\xi}{\sqrt{2^{n+n'+n''+n'''} n! n'!n''! n'''!}}. \nonumber
\end{equation}
$\hat{n}_{\alpha n} =\hat{c}^\dagger_{\alpha n} \hat{c}_{\alpha n}$,
$\Omega_{n} = \Omega_0^B L_{n}(\eta_Z^2) L_0(\eta_Y^2) e^{-(\eta_Y^2 + \eta_Z^2)/2}$,  
and $E_{n}=\hbar \omega_Z(n+1/2)$ are single-particle energies.
In the parameter regime where $ \omega_Z> u I_{n n' n'' n'''}$ (relevant  for current lattice clock  experiments performed at $\mu$K temperatures), to a very good approximation  the leading interaction processes correspond to those ones in which  vibrational quantum numbers are exchanged during the collision,  i.e. $(n=n') \wedge (n''=n''')$ or $(n=n''') \wedge (n''=n') $. Since these  processes conserve     the number of particles per mode, for  an initial state with at most one atom per mode ($g$-polarized state),  it is possible to reduce  $\hat H$  to a  spin-$1/2$ model.  Denoting  $\vec{ n}= \{ n_1,\dots,  n_N\}$ the initially populated modes, the Hamiltonian becomes
\begin{eqnarray}
\hat H^{S}_{\vec{ n}}/\hbar &=& -\delta {S}^z - \sum_{j=1}^N{\Omega_{n_j} {S}^{x}_{n_j}} \nonumber \\ 
&-&\sum_{j=1}^N\sum_{ j'\neq j }^N{\frac{ U_{n_j ,n_{j'}} }{2}(\vec{ S}_{n_j}\cdot \vec{ S}_{ n_{j'}} - 1/4)}. 
\label{manyone}
\end{eqnarray}
Here  $\vec{{S}}_{n_j} = \frac{1}{2}\sum_{\alpha,\alpha'}\hat{c}^\dag_{\alpha n_j}\vec{\sigma}_{\alpha \alpha'}\hat{c}_{\alpha' n_j}$, where $\vec \sigma$ are Pauli matrices in the $\left\{e,g\right\}$ basis, $S^{\tau=x,y,z} = \sum_{j=1}^N {S}^{\tau}_{n_j}$, and constant terms were dropped.
The quantity $U_{n_j, n_{j'}}=u I_{n_j n_j n_{j'} n_{j'}}\equiv u I_{n_j,n_{j'}}$.

The rotational invariance of the interaction term  in $\hat H^S_{\vec{ n}}$ ($\propto U_{n_j,n_{j'}}$) is  key for understanding the basic features of the model.
 Due to the rotational symmetry the interaction term is  diagonal in the collective angular momentum  basis $|S,M,q\rangle$, satisfying $\vec{S}^2|S,M,q\rangle=S(S+1)|S,M,q\rangle$ and $S^z|S,M,q\rangle=M|S,M,q\rangle$, with $S=0 \left(\frac{1}{2}\right),\dots N/2$ and $-S\leq M\leq S$.
Here the extra label $q$ is required to uniquely specify each state. 
 The fully symmetric (Dicke)
$S=N/2$ states 
do not interact. They are unique and the label $q$ can be omitted for them.

To proceed further, we use the fact  that $U_{n_j,n_{j'}}$
is a slowly varying
function of $|n_j-n_{j'}|$. $ I_{n_{j}\gg 0,n_{j'}\gg 0}\to \frac{1}{\pi \sqrt{2|n_j-n_{j'}|}}$.
In the regime ( $k_B T \gg  N \hbar \omega_Z$) where the occupied modes $\vec{ n}$ are sufficiently sparse for the behavior of $U_{n_j,n_{j'}}$ to be dominated by its slowly varying part, we can approximate $U_{n_j,n_{j'}} \to {U}_{\vec{n}}\equiv \sum_{j,j'\neq j} U_{n_j,n_{j'}}/(N(N-1))  $ and
\begin{eqnarray}
\hat H^{S}_{\vec{ n}}/\hbar \approx -   \delta \hat{S}^z - \sum_{j=1}^N{ \Omega_{n_j} \hat{S}^{x}_{n_j}} -  \frac{ {U}_{\vec{n}}}{2} (\vec{\hat S}\cdot \vec{\hat S}). \label{manytwo}
\end{eqnarray}Again constant terms have been dropped.

Under this approximation it is possible to have an analytic treatment of the many-body dynamics.
With this purpose in mind, it is convenient to go to a  rotated  basis and rewrite  the Hamiltonian  as
\begin{eqnarray}
&&\hat{H}^S_{\vec{ n}}/\hbar = \sqrt{\delta^2 +\bar{\Omega}_{\vec{n}}^2} {s}^z \nonumber \\
&&+\sum_{j=1}^N \delta\Omega_{n_j} (\cos \theta {s}^x_{n_j} +\sin\theta {s}^z_{n_j}) -\frac{{U}_{\vec{n}}}{2} \left ( {\vec{s}}\cdot {\vec{s}}\right) \label{manyrot}
\end{eqnarray} 
where ${{s}}^z =\vec{a}\cdot {\vec{S}}$, $\vec{a}=(\sin\theta,0,\cos\theta)$,  $\theta=\arcsin \left (\frac {\bar{\Omega}_{\vec{n}}}{\delta^2+\bar{\Omega}_{\vec{n}}^2}\right)$ , $\delta\Omega_{n_j} =\Omega_{n_j}-\bar{\Omega}_{\vec{n}}$ and
${\bar{\Omega}_{\vec{n}}}=\frac{1}{N}\sum_{j=0}^N\Omega_{n_j}$ the mean Rabi frequency.
We will consider $\delta\Omega_{n_j}$ as our perturbative parameter. To zero order in it, the eigenstates of the Hamiltonian are conveniently described in terms of   angular momentum eigenstates in the rotated basis, $|S,m,k\rangle$ (the quantum number $S$ is conserved in  rotations), satisfying ${\vec{s}}^2|S,m,k\rangle=S(S+1)|S,m,k\rangle$ and ${s}^z=m|S,m,k\rangle$.

At time $t=0$ all the atoms are in the $g$ state and thus initially $S=N/2$. Non-zero $\{\delta \Omega_{n_j}\}$  induce transitions outside the $S=N/2$ manifold. However, to first order in perturbation theory, the term proportional to  $\{\delta \Omega_{n_j}\}$ can only induce transitions to states with $S=N/2-1$ due to its  linear dependence on  $s^{x,z}$. This implies that the knowledge of the eigenstates and eigenvalues within the  $S=N/2, N/2-1$ manifolds is enough to characterize the perturbative dynamics.

The $|N/2,m\rangle$ eigenstates are just the well known Dicke states invariant under particle permutation. They have energies given by $E_{N/2,m}=\hbar \omega_{N/2,m}=\hbar\sqrt{\delta^2 +\bar{\Omega}_{\vec{n}}^2} m $. Since the initially prepared state is a fully polarized state in the old basis, in the rotated basis it corresponds to a superposition of $|N/2,m\rangle$ states  with amplitude probabilities   determined by the Wigner rotation matrices:
\begin{widetext}
\begin{equation}
    |\psi(0)\rangle_{\vec{n}}=|gg \dots g\rangle= \sum_{m=-N/2}^{N/2}
\sqrt {\left(
         \begin{array}{c}
           N \\
           m+N/2 \\
         \end{array}
       \right)
 } \cos^{N/2-m} \left( \frac{\theta}{2}\right) \sin^{N/2+m} \left( \frac{\theta}{2}\right)|N/2,-m\rangle
\end{equation}
\end{widetext}
The states with $S=N/2-1$ are the so called spin-wave states. They can be written in terms of Dicke states as:
\begin{eqnarray}
&&|N/2-1,m,k\rangle=\left(\frac{(N-1)}{(N/2-m+1)(N/2-m)} \right)^{1/2} \nonumber \\
&&\times\sum_{n=1}^{N} e^{i 2\pi k n/N} {s}^{+}_n|N/2,m-1\rangle
\end{eqnarray} 
with $k=1,\dots N-1$. These states have energy $E_{N/2-1,m}/\hbar= \omega_{N/2-1,m}= \sqrt{\delta^2 +\bar{\Omega}_{\vec{n}}^2} m +  \frac{N}{2}  {U}_{\vec{n}}$. From the energy it is clear that the population of these states  will give rise to an interaction energy shift. For  the simple case $N=2$ described in the main text, there is a unique spin wave state which corresponds to the singlet state with energy $ {U}_{\vec{n}}$.

If we write our time evolving many-body state as
\begin{eqnarray}
&&|\psi(t)\rangle_{\vec{n}} = \sum_m c_{m}(t)e^{-i t\omega_{N/2,m}} |N/2,m\rangle \nonumber \\
&&+  \sum_{m,k} b_{m,k}(t)e^{-i t \omega_{N/2-1,m}} |N/2-1,m,k\rangle
\end{eqnarray} 
then the excited state population is given by
 \begin{eqnarray}
N^e_{\vec{n}}(t)&=& \frac{N}{2}+\langle S_z(t)\rangle \nonumber \\
&=& \frac{N}{2}+\cos\theta\langle s_z(t)\rangle- \sin\theta\langle s_x(t)\rangle
\end{eqnarray}

The following transition matrix elements are required for  the perturbative calculations:
\begin{widetext}
\begin{eqnarray}
&&\langle N/2,m |2{s}_n^z | N/2-1,\tilde{m},k\rangle =2 e^{2i \pi k n/N}\sqrt{\frac {(N/2)^2-m^2}{N^2 (N-1)}} \delta_{m,\tilde{m}} \\
&&\langle N/2,m |{s}_n^+ | N/2-1,\tilde{m},k\rangle =- e^{2i \pi k n/N}\sqrt{\frac {(N/2+m)(N/2+ m-1)}{N^2 (N-1)}} \delta_{m,\tilde{m}+1} \\
&&\langle N/2,m |{s}_n^-| N/2-1,\tilde{m},k\rangle =- e^{2i \pi k n/N}\sqrt{\frac {(N/2-m)(N/2- m-1)}{N^2 (N-1)}} \delta_{m,\tilde{m}-1} \\
&&\langle N/2-1,m,k |2{s}_n^z | N/2-1,\tilde{m},\tilde{k}\rangle =(-2 e^{2 i \pi (\tilde{k}-k) n/N}+ N\delta_{k,\tilde{k}})\frac {2m}{N (N-2)} \delta_{m,\tilde{m}} \label{mat1} \\
&&\langle N/2-1,m,k |{s}_n^+ | N/2-1,\tilde{m},\tilde{k}\rangle =(-2 e^{2 i \pi (\tilde{k}-k) n/N}+ N\delta_{k,\tilde{k}})\label{mat2}\\&&\notag
\quad\quad\quad\quad\quad\quad\quad \quad\quad\quad\quad\quad\quad\quad \quad\quad  \times \frac {\sqrt{(N/2+m-1)(N/2-m)}}{N (N-2)} \delta_{m,\tilde{m}+1} \notag \\
 &&  \langle N/2-1,m,k |{s}_n^-| N/2-1,\tilde{m},\tilde{k}\rangle =(-2 e^{2 i \pi (\tilde{k}-k) n/N}+ N\delta_{k,\tilde{k}}) \label{mat3}\\&& \quad\quad\quad\quad\quad\quad\quad \quad\quad\quad\quad\quad\quad\quad\quad\quad   \times \frac {\sqrt{(N/2+m)(N/2-m-1)}}{N (N-2)} \delta_{m,\tilde{m}-1} \notag
\end{eqnarray}
\end{widetext}
The matrix elements in Eqs.(\ref{mat1}-\ref{mat3}) are valid only for $N>2$ and are all zero for $N=2$.

Using those matrix elements one can show after some algebra that  $ N^e_{\vec{n}}(t)$ depends on $\Omega_{n}$ only through the mean Rabi frequency, $\bar{\Omega}_{\vec{n}}$,  and the standard deviation of the Rabi frequency, $\Delta{\Omega}_{\vec{n}}=\sqrt{\sum_{n}\Omega_{n}^2/N-\bar{\Omega}_{\vec{n}}^2}$. More explicitly
\begin{equation}
 N^e_{\vec{n}}(t)=N^{e(0)}_{\vec{n}}(t)+ \Delta\Omega_{\vec{n}}^2 N^{e(2)}_{\vec{n}}(t)+ \mathcal{O}(\Delta\Omega_{\vec{n}}^3)
\end{equation} where the superscript ${}^{(0)}$ indicates a homogeneous excitation.
\begin{equation}
 N^{e{(0)}}_{\vec{n}}(t)= N\frac{\bar{\Omega}_{\vec{n}}^2}{\bar{\Omega}_{\vec{n}}^2+\delta^2} \sin^2\left (\frac{t \sqrt{\bar{\Omega}_{\vec{n}}^2+\delta^2}}{2}\right),\label{hom}
 \end{equation} Note that since  $\langle N/2,m |\hat{s}_{x,y,z} | N/2-1,\tilde{m},k\rangle =0$ then  $N^e_{\vec{n}}(t)$ does  not have first order corrections in $\Delta \Omega_{\vec{n}}$.

\subsubsection{ Analytic evaluation of the clock frequency shift (CFS)}
In clock experiments based on Rabi interrogation the clock frequency shift, CFS,  $\Delta \nu $ is measured  by first locking  the spectroscopy laser  at  two  points, $\delta_{1,2}$, of equal height
in the transition lineshape (equal final excited state fraction under the initial condition of all atoms in state $g$)  and then determining the change in the mean frequency as the interaction parameters  or density  are varied,
$\Delta \nu =(\delta_1+\delta_2)/2$.
To calculate the shift,   we
Taylor expand  $\delta_{1,2}$  around the zero order values, $\pm \delta_1^{(0)}$,  and to lowest nonvanishing order
obtain
\begin{eqnarray}
\Delta \nu_{\vec{n}}  &\approx&
\frac{ N^{e(2)}_{\vec{n}}(t,\delta_1^{(0)})-N^{e(2)}_{\vec{n}}(t,-\delta_1^{(0)})}{2 \frac{\partial N^{e(0)}_{\vec{n}}(t,\delta)}{\partial\delta}|_{\delta_1^{(0)}}} \nonumber \\
&=&\frac{ C_{\vec{n}} }{4 \pi  D_{\vec{n}} }
\label{geshi1}
\end{eqnarray} 
with
\begin{widetext}
\begin{align}
D&=\frac{\delta  \bar{\Omega }^2 \left(-2+2 \cos{\left[t \sqrt{\delta ^2+\bar{\Omega }^2}\right]}+t \sqrt{\delta ^2+\bar{\Omega
}^2} \sin{\left[t \sqrt{\delta ^2+\bar{\Omega }^2}\right]}\right)}{\left(\delta ^2+\bar{\Omega }^2\right)^2}\\
C&=\frac{2   \Delta \Omega
^2 \delta}{N^2 U^2 \left(\delta ^2+\bar{\Omega }^2\right)^3 \left(-N^2 U^2/4+\delta ^2+\bar{\Omega }^2\right)^2} \notag \\
&\times \left (2 N U \bar{\Omega }^4 \left(\delta ^2+\bar{\Omega }^2\right)^2-2 N U \cos[ t N U/2] \bar{\Omega }^2 \left(-N U/2+\bar{\Omega }\right) \left(N U/2+\bar{\Omega
}\right) \left(\delta ^2+\bar{\Omega }^2\right)^2\right. \notag\\
&+ N^5 U^5 \bar{\Omega }^2 /32\left(-7 \delta ^2+2 \bar{\Omega }^2\right)+N^3 U^3/8 \cos\left[t \sqrt{\delta ^2+\bar{\Omega }^2}\right]^2 \bar{\Omega
}^2 \left(-N^2 U^2/4+\delta ^2+\bar{\Omega }^2\right) \left(\delta ^2+2 \bar{\Omega }^2\right) \notag \\
&-N^3 U^3/8 \left(8 \delta ^6+5 \delta ^4 \bar{\Omega }^2+3 \delta ^2 \bar{\Omega }^4+6 \bar{\Omega }^6\right) \notag \\
&+2 N U \cos\left[t \sqrt{\delta ^2+\bar{\Omega }^2}\right] \left(-\cos[ t N U/2] \left(-N U/2+\bar{\Omega }\right) \left(N U/2+\bar{\Omega
}\right) \left(\delta ^2+\bar{\Omega }^2\right)^2 \left(2 \delta ^2+\bar{\Omega }^2\right)\right.\notag \\
&+\left.\bar{\Omega }^2 \left(N^4 U^4 \delta ^2/32+\left(\delta ^2+\bar{\Omega }^2\right)^2 \left(2 \delta ^2+\bar{\Omega }^2\right)-N^2 U^2/4 \left(\delta
^2+\bar{\Omega }^2\right) \left(5 \delta ^2+\bar{\Omega }^2\right)\right)\right)\notag\\
&+\sin\left[t \sqrt{\delta ^2+\bar{\Omega }^2}\right] \left(- t N U \bar{\Omega }^2 \sqrt{\delta ^2+\bar{\Omega }^2} \left(-N^2 U^2/4+\delta
^2+\bar{\Omega }^2\right) \left(N^2 U^2/4 \left(\delta ^2-2 \bar{\Omega }^2\right)\right.\right. \notag\\
&\left.+2 \bar{\Omega }^2 \left(\delta ^2+\bar{\Omega }^2\right)\right)+4 \left(\delta ^2+\bar{\Omega }^2\right)^{5/2} \left(N^4 U^4/16 +\bar{\Omega
}^4+N^2 U^2/4 \left(\delta ^2-2 \bar{\Omega }^2\right)\right) \sin[ t N U/2]\notag\\
&\left.\left.-N^3 U^3/8 \bar{\Omega }^2 \left(-N^2 U^2/4+\delta ^2+\bar{\Omega }^2\right) \left(\delta ^2+2 \bar{\Omega }^2\right) \sin\left[t
\sqrt{\delta ^2+\bar{\Omega }^2}\right]\right)\right).
\end{align}
\end{widetext}
Here we have omitted the subscript ${\vec{n}}$ but it is understood.

So far we have assumed a fixed set of populated modes, $\vec{n}$. At finite temperature, expectation values need to be calculated by averaging  over all possible  combinations of modes $\{ \vec{n}\}$
weighted  according to their  Boltzmann factor:  $\langle \mathcal {O}\rangle_{T_Z} =\frac{\sum_{\vec{n}}\mathcal {O}_{\vec{n}} e^{-E_{\vec{n}}/(k_B T_Z)} }{\sum_{\vec{n}} e^{-E_{\vec{n}}/(k_B T_Z)}}$, with $E_{\vec{n}}=\sum_j E_{n_j}$. The thermally averaged expression of the shift becomes

\begin{eqnarray}
\Delta \nu \equiv \langle \Delta \nu_{\vec{n}} \rangle_{T_Z} &\approx&\frac{\langle C_{\vec{n}} \rangle_{T_Z}}{4 \pi  \langle D_{\vec{n}} \rangle_{T_Z}}
\label{geshi2}
\end{eqnarray} Grouping all temperature dependent terms  in a temperature dependent coefficient,  we obtain  the following scaling behavior of the shift: $\Delta \nu \propto A(T_Z,N) N u \eta_z^4 $ in the weakly interacting regime consistent with prior mean field analysis \cite{Oktel1999app, Oktel2002app}
  and $ \Delta \nu  \propto B(T_Z,N)(\Omega_0^B)^2 \eta_z^4/  (N u)$ in the strongly interacting regime.

\subsection{Estimation of the tunneling rates under current experimental conditions}

\subsubsection{Single Particle tunneling}

Here we discuss the  tunneling rates in the weakest lattice potential which is along the $Y$ direction. The  trap frequency is $\omega_Y/( 2 \pi)=55$ kHz which corresponds to a lattice depth of  $V_Y=64 E_r$ with $ E_r=\hbar^2 k_L^2/(2m_\mathrm{Sr})$ the photon recoil energy and $m_\mathrm{Sr}$ the atomic mass.

\begin{figure}
     \begin{center}
        \includegraphics[width=0.9\columnwidth]{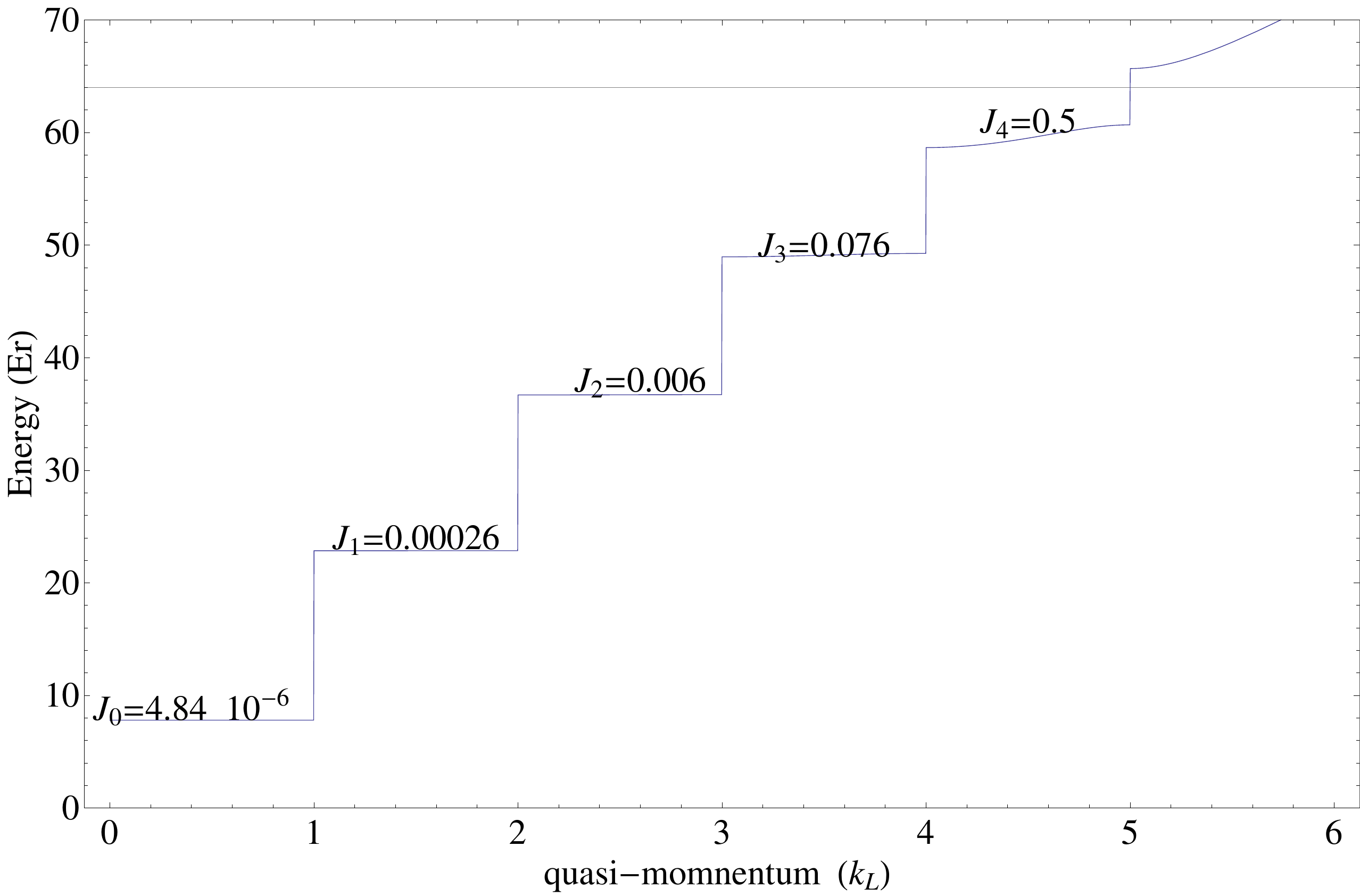}
        \caption{Dispersion relation for atoms in a $64 E_r$ lattice and the corresponding tunneling energies (in recoil energy) of the relevant bands. The solid grid line indicates the lattice depth.}\label{bandstr}
    \end{center}
\end{figure}

The energy levels  of a sinusoidal lattice  potential, $V_Y \sin^2(k_L Y)$,  admit analytic solutions in terms of Mathieu functions and their characteristic values.  We  use those to compute the band structure. In Fig.~\ref{bandstr} we plot the dispersion relation showing the five relevant bound states. In addition we write the corresponding tunneling energy, $J$ for each  band  in units of $E_r$. We evaluate the thermal average of the tunneling energy using  a Boltzmann distribution and obtain $\langle J\rangle_{T_Y}=(0.007, 0.024)E_r$ for experimental temperatures of   $T_Y\sim (2.5,4)\mu$K respectively which correspond to tunneling times of $(38,11)$ ms.

In the experiment in addition to the lattice potential along $Y$, the atoms feel the  dipole potential generated by the $X$ lattice   as well as the gravitational  force. The total  external potential along the $Y$ direction is approximately given by:
\begin{eqnarray}
  V_t(Y) &=& V_Y \sin^2(k_L Y) \nonumber \\
  &+&\frac{1}{2} m_\mathrm{Sr} \omega_{DY}^{2}Y^2 + m_\mathrm{Sr} g Y\\
  V_t(Y) &=& V_Y \sin^2(k_L Y) \nonumber \\
  &+&\frac{1}{2} m_\mathrm{Sr}\omega_{DY}^{2}\left(Y+\frac{g}{m_\mathrm{Sr} \omega_{DY}^{2}}\right)^2 \nonumber \\
&-&\frac{g^2}{2m_\mathrm{Sr} \omega_{DY}^{2}}
\end{eqnarray}
   with $\omega_{DY}= 4 V_X/(m_\mathrm{Sr} w_0^2)\sim 2 \pi \times 485 $ Hz and $w_0\sim 32 \mu$m the beam waist.
   The additional parabolic potential creates an energy offset between adjacent lattice sites which tends to suppress tunneling.
   For a mean energy offset $\Delta$ between two adjacent lattice sites the effective tunneling rate becomes $J_{eff}=J^2/\sqrt{J^2+\Delta^2}$.
   Due to the harmonic nature of the confinement $\Delta$  depends on the actual position in the lattice.
    For two  adjacent lattice sites, at $Y_1= j a$ and $Y_2= (j+1) a$ respectively, the  mean  energy offset is $\Delta_j= (V_t(Y_2)-V_t(Y_1))/2=
   \frac{1}{2} m_\mathrm{Sr} \omega_{DY}^{2} a^2 (j+1/2)$ with $a$ the lattice spacing.

  As mentioned in the main text the loading procedure of the atoms is such that 100 ìrowsî of tubes are approximately uniformly distributed along $Y$.
  This is consistent with  an average $\bar{ j}=25$, i.e $\bar{\Delta}=\Delta_{25}$.  Using this value  we   estimate $\langle J_{eff}\rangle_{T_Y}=(0.002, 0.007)E_r$ for experimental temperatures of   $T_Y\sim (2.5,4)\mu$K respectively which corresponds to effective tunneling times of $(150,40)$ ms.

  These results demonstrate that for a typical  interrogation time of the order of $80$ ms,  tunneling events  can be neglected, especially for the  $T_Y\sim 2.5\mu$K situation which was the  temperature at which most data points were taken (see main text).

\subsubsection{Interaction assisted tunneling}

  Interactions can modify the effective tunneling rates. However, in typical cold atomic systems interacting via $s$-wave collisions these effects are
  so weak that  can be generally ignored. The role  of interaction assisted tunneling effects  has been  explicitly investigated  in a recent experiment~\cite{uno}. This experiment has been the only one capable of observing super-exchange interactions at energy scales as low as 5 Hz. Nevertheless, regardless of  the high degree of experimental control and precision,  the experimental data was not able to observe any deviation of the single-particle  tunneling due to interactions  compatible with the uncertainties of the lattice depths.

  Furthermore, recent theoretical efforts have predicted that only close to a Feshbach resonance, interaction assisted tunneling terms can start to play a role, see Ref\cite{dos,tres}. This happens when the mean interaction energy of the gas becomes larger than the vibrational energy spacing.

    In our experiment  the condition  $\langle  U \rangle _T\ll \omega_Y $ is very well satisfied  since $\langle  U \rangle _T<$kHz and   $\omega_Y\sim 2 \pi \times 55 $ kHz. Consequently  we can safely neglect interaction assisted tunneling effects.

\subsection{Data analysis methods}

In the experiment described above, the density shift was determined by measuring a differential shift of the Sr clock transition as the number of atoms loaded into the 2D lattice was varied. The sample density was alternated between a high and a low value every two cycles of the experiment, as two interrogations of the clock transition were required to obtain one estimate of the transition center frequency.
The sequential frequency data points are indexed by $l$, which runs from $l=1$ to $l=M$, where $M$ is the total number of frequency points in a data run.
The correlation between the clock transition frequency and the sample density was determined by analyzing consecutive groups (``strings") of frequency data, of length $n$.
If for all odd $l$ the density was at the low value, and for all even $l$ the density was high, the correlated frequency shift is
\begin{equation}
\delta\nu=\frac{1}{M}\frac{1}{2^{n-2}}\sum_{l=1}^M\sum_{m=l}^{n} (-1)^{(m)} \frac{(n-1)!}{(n-m)!(m-1)!}\ \nu_{m}.
\end{equation}
We then perform the same string analysis on the associated sample densities and normalize each string point to obtain the collisional frequency shift per unit density.
A similar analysis was introduced in \cite{Dress1977}, and this method is equivalent to removing a polynomial of order $n-2$ from each $n$-point sequence of data. For a string of length $n=2$, this simply corresponds to isolating the differential shift between the high and low density data, while for a string of length $n=3$, a linear drift is removed. As mentioned in the main text, we used strings of length $n=4$ to analyze our data, thereby eliminating sensitivity to both linear and quadratic drifts. This drift removal is necessary because of the residual drift in the ultrastable Fabry-Perot cavity that served as our frequency reference.

For each data run, we report the mean and standard error of the string points.
As the data strings overlap with one another, they are not completely independent, and a correction factor
\begin{equation}
f_{cor} = \frac{2^{n-1}}{\sqrt{\sum_{m=1}^{n}\left(\frac{(n)!}{(n-m)!(m+1)!}\right)}},
\end{equation}
must be applied to the standard error of the string points. Stated in other words, without this factor there are ``too many points" and dividing the standard deviation of the data set by the square root of the number of measurements will underestimate the error.

A single data run consisted of a period of continuous lock to the clock transition, and therefore there was a significant variation in the lengths of individual runs. This variation in run length is partly responsible for the variation in the errors of the data runs. We applied a cut at the start of each data run to allow sufficient time for the clock laser servo to acquire a stable lock to the atomic transition.

All data runs corresponding to a given set of of experimental conditions were combined to give an estimate of value and the uncertainty of the frequency shift (see Fig.~\ref{fig5} in the main text). The shift value was obtained by taking the error-weighted mean of all the data runs, and the uncertainty was estimated from the variance of the weighted mean. We calculated a $\chi^2$ per degree of freedom, or a reduced $\chi^2$ ($\chi^2_\mathrm{red}$) for each set of data runs, and scaled the final uncertainty by $\sqrt{\chi^2_\mathrm{red}}$.

We analyzed the data using several different protocols, each of which give results that are consistent within the stated uncertainties. Below, we briefly describe each protocol and present the shift each method calculates for the data records presented in Fig.~\ref{fig5} of the main text:

\begin{itemize}
\item considering all of the data as one single unit, without regard to unlocks or different days on which the data was taken. In this case, the central value of the result is taken to be the simple mean of the data, and the uncertainty is estimated by the standard error scaled by a factor to account for the overlap of the string points. For this method we find $(-0.6 \pm 2.6)\e{-17}$ for the 3.5 $\mu$K data, and $(5.0\pm1.7)\e{-17}$ for the 7 $\mu$K data.
\item separating the data into bins of a set number of string points and calculating the simple mean of each bin.  The error bars for each bin are determined from the standard error of the bin and scaled by the correction factor.  This data is then used to calculate the weighted mean and error of the weighted mean for the entire data record.  A reduced chi-square statistic was also calculated and the bin size is adjusted to achieve a reduced chi-square of $1.0$ for each data record.  This method has the advantage that it weights data according to actual scatter of the measurement points so that data with greater signal to noise is weighted more heavily.  With this method, we are also able to consistently compare our longer data records to data records where the data was acquired over very few continuous locks to atoms and even fewer days (making the two following protocols
    inappropriate).  In an effort to present all of our data using a consistent method of analysis, all data point in Fig.~\ref{fig4} of the main text are calculated using this method.  When analyzed with this method, the data records give $(1.2 \pm 1.9)\e{-17}$ for the 3.5 $\mu$K data, and $(5.8\pm1.3)\e{-17}$ for the 7 $\mu$K data (both $\chi^2_\mathrm{red}=1.0$).
\item breaking the data into data runs by day (where each day of data-taking includes multiple segments of continuous lock). Error bars for each data run are determined from the standard error of the run measurements and scaled by the correction factor. Simple means of each data run were used to determine the central value of the shift for that run. The weighted mean and the error of the weighted mean were determined from these data, and a reduced chi-square statistic was calculated. We then scale the error by the square-root of the chi-square value.  For this method the data records give $(0.2\pm2.3)\e{-17}$ with a $\sqrt{\chi^2_\mathrm{red}}=0.78$ for the 3.5 $\mu$K data, and $(5.4\pm1.6)\e{-17}$ with a $\sqrt{\chi^2_\mathrm{red}}=1.1$ for the 7 $\mu$K data (error unscaled).
\item similar to the previous method, but data runs are taken to be periods of continuous lock to the atoms rather than days of data taking. This is the method used in Fig.~\ref{fig5} of the main body of the paper. For this method we find $(0.5\pm2.3)\e{-17}$ with a $\sqrt{\chi^2_\mathrm{red}}=0.73$ for the 3.5 $\mu$K data, and $(5.6\pm1.6)\e{-17}$ with a $\sqrt{\chi^2_\mathrm{red}}=0.84$ for the 7 $\mu$K data (error unscaled).
\end{itemize}

\subsection{Characterization of the optical trap}
We used several techniques to characterize important experimental parameters such as the sample temperature along the strongly- and weakly-confined directions in the 2D lattice, the Lamb-Dicke parameter, and the trapping frequencies. The temperature along the weakly-confined direction ($\hat Z$) was established by fitting the width of the Doppler broadened spectroscopic feature obtained when the clock laser propagated along $\hat Z$. Temperatures along the strongly-confined directions were determined via sideband spectroscopy, as described in \cite{Blatt2009} and the main text.

The trap frequencies along $\hat X$ and $\hat Y$ both vary as the horizontal lattice power is varied. This occurs because the two lattice beams are not orthogonal but instead cross at an angle of $\sim 71^\circ$. The eigenaxes of the resulting 2D harmonic potential therefore are a function of the power in both of the two lattice beams. This effect was properly taken into account when the trap confinement and transverse Lamb-Dicke parameter were determined.

\end{document}